\documentclass[12pt]{amsart}

\usepackage[all, cmtip]{xy}

\usepackage{graphicx, subfig}
\usepackage{epsfig}
\usepackage{amscd}
\usepackage[mathscr]{eucal}
\usepackage{amssymb}
\usepackage{amsxtra}
\usepackage{amsmath}
\usepackage{latexsym}
\usepackage[all]{xy}
\usepackage{enumerate}
\usepackage{mathptmx}    
\usepackage{color}
\usepackage{rotating}
\usepackage[colorlinks=true,linkcolor=red,urlcolor=green,citecolor=blue]{hyperref}
\usepackage[hyperpageref]{backref}
\usepackage{rotating}

\usepackage{color}
\theoremstyle{plain}

\theoremstyle{definition}

\theoremstyle{remark}

\theoremstyle{plain}

\begin{document}

\title[]{Two-dimensional superintegrable systems from operator algebras in one dimension}
\date{\today}

\author[]{Ian Marquette, Masoumeh Sajedi, Pavel Winternitz}

\address{Ian Marquette, School of Mathematics and Physics\\
 The University of Queensland, Brisbane, QLD 4072, Australia} \email{i.marquette@uq.edu.au}
\address{Masoumeh Sajedi, D\'epartement de math\'ematiques et de statistiques\\
 Universit\'e de Montr\'eal, C.P.6128 succ. Centre-Ville, Montr\'eal (QC) H3C 3J7, Canada} \email{sajedim@dms.umontreal.ca}
\address{Pavel Winternitz, Centre de recherches math\'ematiques, D\'epartement de math\'ematiques et de statistiques\\
 Universit\'e de Montr\'eal, C.P.6128 succ. Centre-Ville, Montr\'eal (QC) H3C 3J7, Canada} \email{wintern@crm.umontreal.ca}
  
%
 \maketitle

\begin{abstract}
We develop new constructions of 2D classical and quantum superintegrable Hamiltonians 
allowing separation of variables in Cartesian coordinates. In classical mechanics we start from two functions on a one-dimensional phase space, a natural Hamiltonian $H$ and a polynomial of order $N$ in the momentum $p.$ We assume that their Poisson commutator $\{H,K\}$ vanishes, is a constant, a constant times $H$, or a constant times $K$. In the quantum case $H$ and $K$ are operators and their Lie commutator has one of the above properties. We use two copies of such $(H,K)$ pairs to generate two-dimensional superintegrable systems in the Euclidean space $E_2$, allowing the separation of variables in Cartesian coordinates. All known separable superintegrable systems in $E_2$ can be obtained in this manner and we obtain new ones for $N=4.$

\end{abstract}

\vspace{0.5cm}


 
\newpage

\section{INTRODUCTION}
This article is part of a general study of superintegrable systems in quantum and classical mechanics. In a nutshell a superintgerable system with $n$ degrees of freedom is a Hamiltonian system with $n$ integrals of motion $X_1,...,X_n$ (including the Hamiltonian $H$) in involution and $k$ further integrals $Y_k, 1 \leq k \leq 2n-1.$ The additional integrals $Y_k$ commute (or Poisson commute) with the Hamiltonian, but not necessarily with each other, nor with the integrals $X_i.$ All the integrals are assumed to be well defined functionally independent functions on phase space in classical mechanics. In quantum mechanics they are Hermitian operators in the enveloping algebra of the Heisenberg algebra $H_n$ (or some generalization of the enveloping algebra) and are polynomially independent. For reviews we refer to \cite{kalnins18, Miller:review}.\\
The best known superintegrable systems (in $n$ dimensions) are the Kepler-Coulomb system \cite{Bargmann:Ba, Fock:Fo, pauli26} and the harmonic oscillator \cite{hill40, Moshinsky:HO} with the potentials $\frac{\alpha}{r}$ and $\omega r^2$, respectively.\\
A systematic search for superintegrable systems in Euclidean spaces $E_n$ was started in $1965$ \cite{Fris:Fri, Makanov:Nonrel, Evans:CM, Evans:W, Evans:SW}. The integrals of motion were postulated to be second order polynomials in the momenta with coefficients that were smooth functions of the coordinates. Second order integrals were shown to be related to multiseparation of variables in the Schr\"{o}dinger or Hamilton-Jacobi equation. Integrable and superintegrable systems with integrals that are higher order polynomials in the momenta were considered in \cite{ismail18, calzada14a, calzada14b, demir02, adrian17, adrian18, adrian18-nth, gravel04, gravel02, kalnins11, kalnins12, Snobl:15, marquette07, marquette09a, marquette09b, marquette09c, sajedi17, nikitin04, popperi12, riglioni15, post10, post11, post15, Ranada:Drach, sheftel01, TWT:exact, Tremblay:09, tremblay10, Tsiganov:Drach}.\\
A subset of the articles quoted above was devoted to a search for superintegrable systems in $E_2$ with $2$ integrals of order $2$ and one of order $N$ with $3 \leq N \leq 5.$ The order $2$ ones were the Hamiltonian $H$, the second order one $X$ was chosen so as to ensure separation of variables in Cartesian or polar coordinates, respectively. The third integral $Y$ was of order $N \geq 3$. It turned out that the complexity of the calculations rapidly increased as $N$ increased and that the obvious systematic and straight forward method became impractical for $N > 5$. On the other hand for $N \geq 3$ it turned out that quantum integrable and superintegrable systems could have different potentials than classical ones. In particular quantum superintegrable systems allowed the existence of "exotic potentials" expressed in terms of elliptic functions , Painlev\'e transcendents and general functions having the Painlev\'e property.\\
The purpose of this paper is to further develop and apply a different method of constructing superintegrable systems in two and more dimensions. Namely, we shall study two copies of operator algebras in one dimension, expressed in terms of the coordinates $x$ and $y$, respectively and combine these two to form superintegrable systems in $E_2$. The generalization to $n$ copies and to superintegrable systems in $E_n$ is immediate.\\
The article is organized as follows. In Section $2$ we formulate the problem and show how algebras of operators or functions in one dimension can be used to construct superintegrable systems in two dimensions. This is done both for quantum and classical mechanics. Section $3$ is devoted to the classification of operator algebras in one-dimensional quantum mechanics for operators $K$ of order $1\leq M \leq 5.$ The same problem in classical mechanics, where $H$ and $K$ are functions on a two dimensional phase space is solved in Section $4$. The superintegrable classical and quantum systems in $E_2$ are presented in Section $5$. Section $6$ is devoted to conclusions and a summary of results.\\
\section{The general method}

Let us consider a Hamiltonian in a one dimensional Euclidean space $E_1$

\begin{equation}\label{H1}
H_{1}=\frac{p_{x}^{2}}{2}+V(x)
\end{equation}
where $x$ is a space coordinate. In classical mechanics $p_x$ is the momentum canonically conjugate to $x$ and in quantum mechanics we have $p_x=-i \hbar \partial_x.$
\newline
Let us also consider the polynomial
\begin{equation}\label{Lx}
K_1=\sum_{j=0}^Mf_j(x)p_x^j,
\end{equation}
where $f_M(x) \neq 0,$ and $f_j(x)$ are locally smooth functions.
\newline
Both in quantum and in classical mechanics we can consider the Lie algebra
\begin{equation}\label{lie}
[H_1,K_1]=\alpha K_1 +\beta H_1+\gamma 1, \;[H_1,1]=[K_1,1]=0, 
\end{equation}
where $[.,.]$ is the Lie commutator , or the Poisson commutator, respectively and $\alpha, \beta $ and $\gamma$ are constants. By change of basis we can reduce the algebra (\ref{lie}) into one of the $4$ following forms for $\alpha=\beta=\gamma=0;\; \alpha=\beta=0,\gamma \neq 0;\; \alpha=\gamma=0,\beta \neq 0;$ and $\alpha \neq 0,$ respectively 

\begin{subequations}\label{types}
\begin{equation}
[H_1,K_1]=0,
\end{equation}
\begin{equation}\label{typesb}
[H_1,K_1]=\alpha_1,
\end{equation}
\begin{equation}\label{typesc}
[H_1,K_1]=\alpha_1 H_1,
\end{equation}
\begin{equation}\label{typesd}
[H_1,K_1]=- \alpha_1 K_1,\; \alpha_1 \in \mathbb{R} \backslash 0
\end{equation}
\end{subequations}
where $\alpha_1 \neq 0$ is a constant. This constant could be normalized to $\alpha_1=1$. We however leave it general and use it as a parameter to be chosen later. We shall refer to these relations as Abelian type (a), Heisenberg type (b), conformal type (c), and ladder type (d), respectively.\\
We shall call the systems $\{H_1,K_1\}$ in one dimension "algebraic Hamiltonian systems". The classical case (d) of ladder and the corresponding Hamiltonian and functions $K_1$ that are polynomials of order $3$ and $4$ in momentum have been studied in the references \cite{marquette10b, marquette11, marquette12}. For earlier work on ladder operators and separation of variables see \cite{BM1974}. The case of order 3 of these relations has been discussed in \cite{fushchych97}. Some of these cases have been investigated e.g. in the case (c) \cite{doebner99} and case (b) \cite{Gungor:14}. The quantum case (d) has been studied for particular examples related with fourth and fifth Painlev\'e transcendents \cite{marquette10a, marquette09b, marquette11, veselov93,  andrianov00, carballo04}. Superintegrable deformations of the harmonic oscillator and the singular oscillator and many types of ladder operators have been studied \cite{marquette13a, marquette13c, marquette14}. The Heisenberg type relations have been investigated in a recent paper \cite{Gungor:14}. The Abelian type (a) has been studied by Hietarinta for third order operators and was referred to as pure quantum integrability \cite{hietarinta89,hietarinta98}. Furthermore, for the case (a) some interesting algebraic relations have been discussed \cite{turbiner94,turbiner96}.
\newline
The existence of such operators $K_1$ will impose constraints on the potential $V(x)$ and on the coefficients $f_j(x)$ in the polynomial $K_1.$ We shall construct such systems proceeding by order $M$ in the following sections, both in quantum and classical physics.
\newline
We consider a second copy of $E_1$ with the corresponding Hamiltonian $H_2$ and operator $K_2$ satisfying one of the relations (a),(b),(c),(d) with
\begin{equation}\label{Ly}
K_2=\sum_{j=0}^Ng_j(y)p_y^j.
\end{equation}
Now let us consider the two-dimensional Euclidean space $E_2$ with the Hamiltonian
\begin{equation}\label{H}
H=H_1+H_2=\frac{1}{2}(p_x^2+p_y^2)+V_1(x)+V_2(y).
\end{equation}
This Hamiltonian is obviously integrable because it allows the separation of variables in Cartesian coordinates, i.e. it allows an independent second order integral
\begin{equation}\label{A}
A=K_1-K_2,
\end{equation}
where $K_1$ and $K_2$ are second order Abelian type operators.\\
We will use operators $K_1, K_2$ of (\ref{Lx}) and (\ref{Ly}) to generate integrals of motion $K$ in $E_2.$ Below the notation is $(u,v)$. The first label applies to the $x$ axis and the second one to the $y$ axis. Both $u$ and $v$ take the values $a,b,c$ and $d$, depending on the type of algebraic Hamiltonian system in (\ref{types}). The combinations that lead to superintegrable systems in $E_2$ are\\
\textbf{I.(a,a):}\\
Obviously, any linear combination
\begin{equation}
K=c_1K_1+c_2 K_2
\end{equation}
satisfies $[H,K]=0$ and is hence an integral of motion. The interesting point is that in the case $(a)$ , $H_1$ and $K_1$ in $E_1$ can not be polynomially independent, however in $E_2$ the operators $H,$ and $K$ can be. The case of (\ref{H}) and (\ref{A}) is a trivial example. For higher order operators we shall produce nontrivial examples below (in quantum mechanics).
\newline
\textbf{II.(b,b):}\\
The operator
\begin{equation}
K=\alpha_2 K_1-\alpha_1 K_2
\end{equation}
will commute with $H.$
\newline
\textbf{III.(c,b):}\\
An integral of motion is $K=\alpha_2 K_1-\alpha_1 H_1 K_2.$
\newline
\textbf{IV.(d,d):}\\
The case (d) is somewhat more complicated. We change notations slightly and introduce an operator $K_1^{\dagger} $ adjoint to $K_1.$
\begin{equation}
K_1^- \equiv K_1,\; K_1^{\dagger} =(K_1^-)^{\dagger}.
\end{equation}
We now have
\begin{align}\label{KKd}
&[H_1,K_1^-]=-\alpha_1 K_1^- \nonumber\\
&[H_1,K_1^{\dagger}]=\alpha_1 K_1^{\dagger} \nonumber\\
&K_1^{\dagger} K_1^- =\sum_{n=0}^{k_x} a_n H_1^n
\end{align}
The fact that $K_1^\dagger K_1^-$ is a polynomial in $H_1$ follows from the commutation relation $[H_1,K_1^\dagger K_1^-]=0$ \cite{BC1923}.
\newline
The same relations are introduced for $H_2, K_2^-$ and $K_2^\dagger$. In $E_2$ we have
\begin{equation}
[H_1+H_2, (K_1^\dagger)^m(K_2^-)^n]=(m\alpha_1-n\alpha_2)(K_1^\dagger)^m(K_2^-)^n
\end{equation}
To obtain an integral of motion we impose a rationality constraint on $\alpha_1$ and $\alpha_2,$ namely
\begin{equation}
\frac{\alpha_1}{\alpha_2}=\frac{n}{m}.
\end{equation}
With this constraint $(K_1^\dagger)^m(K_2^-)^n$ are all integrals of motion and
\begin{equation}
K=(K_1^\dagger)^m(K_2^-)^n-(K_1^-)^m(K_2^\dagger)^n.
\end{equation}
with $m$ and $n$ mutually prime $K$ is the lowest order polynomial amongst them.
\newline
\textbf{V.(c,c):}\\
The integral of motion in this case is
\begin{equation}
K=\alpha_2 H_2K_1-\alpha_1 H_1K_2
\end{equation} 
\newline
Other possible combinations are
\newline
\textbf{VI.(a,d):}\\
An integral of motion is $K_1+K_2^{-}K_2^{\dagger}$. However we have $K_2^\dagger K_2^-=P(H_2),$ so this integral is trivial ( a polynomial in $H_1$ and $H_2$).

\begin{table}[h]\label{Tint}
\begin{tabular}{|l|l|l|l|l|}
  \hline
 Case  &  Type &  Integral type & K  &  Order of $K$  \\[0.05cm]
  \hline	
 1  &  (a,a) & polynomial  &  $K_1+K_2$ &  $max(k_{x},k_{y})$    \\[0.05cm]
  \hline
 2  &  (b,b)  & polynomial & $\alpha_{2} K_1-\alpha_{1}K_2$ & $max(k_{1},k_{2})$  \\[0.05cm]
  \hline
 3  &  (c,b) & polynomial &  $\alpha_{2}K_1-\alpha_{1}H_{1}K_2$ & $max(k_{1},k_{2}+2)$      \\[0.05cm]
  \hline
 4  &  (d,d) & polynomial &  $(K_{1}^{\dagger})^{m} (K_2^{-})^{n} - (K_1^{-})^{m} (K_2^{\dagger})^{n} $ & $(m k_{1}+n k_{2}-1)$     \\[0.05cm]
  \hline	
 5  &  (c,c) & polynomial &  $\alpha_{2} H_{2}K_1-\alpha_{1} H_{1}K_2$ & $max(k_{1}+2,k_{2}+2)$    \\[0.05cm]
  \hline
 6  &  (a,d) & polynomial &  $K_1-K_2^{-}K_2^{\dagger}$ & $max(k_{1},2k_{2})$   \\[0.05cm]
  \hline		
 7  &  (b,d) & non polynomial &  $e^{i\frac{\alpha_{2}}{\alpha_{1}}K_1}K_2^{-}$ &  -   \\[0.05cm]
  \hline			
 8  &  (c,d) & non polynomial &  $e^{i\frac{\alpha_{2}}{\alpha_{1}}K_1}K_2^{H_{1}}$ &  -      \\[0.05cm]
 
  \hline							
						
\end{tabular}
\caption{\footnotesize{Integrals of motion in $E_2$}}
\end{table}

In Table $1$, $k_1=order(K_1)$ and $k_2=order(K_2).$ For the operator of type (d), setting $K_1=K_1^{-}$, we have $[H_1,K_1^{\pm}]=\pm \alpha _1K_1^{\pm}.$ Also in the case 4, $m\alpha_1=n\alpha_2=\lambda$.
\newline
Let us consider $A$ as the second order integral of motion introduced in (\ref{A}) and $B$ as the $M$th order one. In the classical case, the polynomial Poisson algebra $\mathcal{P}_M$, generated by functions $A$ and $B$ has Poisson brackets given by 
\newline
\begin{equation}
\{A,B\}_{p}=C, \{A,C\}_{p}=R(A,B,H), \{B,C\}_{p}=S(A,B,H).
\end{equation}
\newline
The polynomial Lie algebra, $\mathcal{ L}_M$, which is the $M$th order analogue of the classical Poisson
algebra $\mathcal{P}_M$, has bracket operation given by
\begin{equation}
[A,B]=C, [A,C]=\tilde{R}(A,B,H), [B,C]=\tilde{S}(A,B,H)
\end{equation}

with further constraints on parameters from the Jacobi identity. Further information on the algebra is given in Table $2$.\\
In this article we pursue the case where $B$ is a polynomial. The cases $7$ and $8$ of Table $1$ will be treated elsewhere.

\begin{table}\label{Talg}
\begin{center}
\begin{tabular}{|l|l|l|l|l|l|}
  \hline
 Case  &  Type &  $R(A,B,H)$ & $S(A,B,H)$  & $\tilde{R}(A,B,H)$   &  $\tilde{S}(A,B,H)$   \\[0.05cm]
  \hline	
 1  &  (a,a) & 0  &  0 &  0  & 0     \\[0.05cm]
  \hline
 2  &  (b,b)  & 0 & 0 & 0  &  0    \\[0.05cm]
  \hline
 3  &  (c,b) & 0 & $\kappa(H+A)$  & 0  & $\kappa(H+A)$    \\[0.05cm]
  \hline
 4  &  (d,d) & $-4 \lambda^2 B$ &  $T(A,H)$   & $4\lambda ^2 B$  & $\tilde{T}(A,H)$  \\[0.05cm]
  \hline	
 5  &  (c,c) & 0 &  $\frac{\kappa}{2}A(H^2-A^2) $  & 0  & $\frac{\kappa}{2}A(H^2-A^2)$ \\[0.05cm]
  \hline
 6  &  (a,d) & 0 &  0  & 0 & 0 \\[0.05cm]
  \hline		
 7  &  (b,d) & $-4\alpha_{2}^2 B$ & 0  &  -  &  - \\[0.05cm]
  \hline			
 8  &  (c,d) & $-4\alpha_{2} B$ &  0 &  -  &  -  \\[0.05cm]
  \hline

\end{tabular}
\caption{\footnotesize{Polynomial algebra}}
\end{center}
\end{table}

\newpage
In Table $2,$ $\kappa=\alpha_{1}^2\alpha_{2}^2$ and

$$T(A,H)=4\lambda P(\frac{H+A}{2})^{m-1} P(\frac{H-A}{2})^{n-1}\big(n^{2}Q(\frac{H+A}{2})P(\frac{H-A}{2})-m^{2}Q(\frac{H-A}{2})P(\frac{H+A}{2})\big),$$

\[\tilde{T}(A,H)=-2\prod_{i=1}^{m}Q(\frac{H}{2}+\frac{A}{2}-(m-i)\alpha_{1})\prod_{j=1}^{n}S(\frac{H}{2}-\frac{A}{2}+j\alpha_{2})\]
\newline
with $K_1^+ K_1^-=P(H_1)$, and $\{K_1^-,K_1^+\}=Q(H_1)$.

\section{Classification of quantum algebraic systems in one dimension}

We consider the one dimensional Hamiltonian (\ref{H1})
and the $M$th order operator $K_1$ (\ref{Lx}) and their commutator $[H_1,K_1]$.\\
Once $[H_1,K_1]$ is chosen to be equal to $0, \alpha_1, \alpha_1H_1$ or $-\alpha_1K_1$ as in (\ref{types}), this will provide us with determining equations for the potential $V(x)$ and the coefficients $f_j(x), \; 0\leq j \leq M$ in the operator $K_1.$\\
Using $p_x=-i \hbar \partial_x \equiv -i \hbar D$ we obtain the following operator of order $M+1$.
\begin{align}\label{comutD}
[H_1,K_1]=-\frac{\hbar^{2}}{2}\sum_{l=0}^M (-i\hbar)^{l}( f_{l}'' D^l + 2 f_{l}' D^{l+1}) -\sum_{l=1}^{M}(-i\hbar)^{l}f_l\sum_{j=0}^{l-1} C_j^l V^{(l-j)}D^{j} 
\end{align}
where $C_j^k$ are the Newton binomial coefficients.\\
In order to obtain the determining equations for arbitrary $M$ we must reorder the double summation in the second term in (\ref{comutD}). We obtain
\begin{align}\label{comutDg}
[H_1,K_1]=\sum_{l=0}^{M+1}Z_lD^l
\end{align}
with
\begin{subequations}\label{comutz}
\begin{equation}\label{comutz1}
Z_{M+1}=(-i\hbar )^{M+2} f_M',
\end{equation}
\begin{equation}\label{comutz2}
Z_{M}=-\frac{\hbar ^2}{2}(-i \hbar)^{M-1}(2f_{M-1}'-i\hbar f_{M}''),
\end{equation}
\begin{equation}\label{comutz3}
Z_{l}=-\frac{\hbar ^2}{2}(-i \hbar)^{l-1}(2f_{l-1}'-i\hbar f_{l}'')-\sum_{j=l+1}^{M}(-i\hbar)^{j}f_j C_l^j V^{(j-l)},\; 1 \leq l \leq M-1,
\end{equation}
\begin{equation}\label{comutz4}
Z_0=-\frac{\hbar ^2}{2}f_0''-\sum_{j=1}^{M} (-i\hbar)^{j} f_j V^{(j)}
\end{equation}
\end{subequations}

The determining equations for arbitrary $M\geq 1$ are as follows\\
\textbf{Case(a):}\\
We have 
$$Z_l=0,\; 0\leq l \leq M+1.$$
In particular equations (\ref{comutz1}) and (\ref{comutz2}) imply $f'_M=0, f'_{M-1}=0.$ Equations (\ref{comutz3}) provide expressions for $f_l''$ in terms of the potential $V(x)$ and its derivatives for $l=0,...,M-1$. Substituting $f_l$ into (\ref{comutz4}) we obtain a nonlinear ODE for the potential $V(x),$\\
\textbf{Case(b):}\\
The determining equations are
\begin{align}
&Z_0=\alpha_1,\; Z_l=0, 1 \leq l \leq M+1.\\\nonumber
\end{align}
Hence the functions $f_l,\;0 \leq l \leq M$ are the same as in case (a) but the equation for the potential $V(x)$ is modified.\\
\textbf{Case(c):}\\
This case arises for $M \geq 2.$ The determining equations are
\begin{align}
&Z_l=0, l \neq 0,2\\\nonumber
&Z_0=\alpha_1 V(x), \quad Z_2=-\alpha_1 \frac{\hbar}{2}.
\end{align}
Again, equation (\ref{comutz4}) provides an ODE for $V(x)$.\\
\textbf{Case(d):}\\
The determining equations are

\begin{align}
&Z_{M+1}=0,\quad Z_l=\alpha_1 f_l,\quad 0 \leq l \leq M.
\end{align}
In this article we concentrate on the cases $1 \leq M \leq 5$ but it is clear that one can proceed iteratively for any given $M.$\\
Let us now solve the determining equations for $1 \leq M \leq 5$.\\
The notation used below is $V_{\gamma_M}$ where $\gamma=a,b,c,d$ refers to the four different cases in equation (\ref{types}) and $M=1,2,...,5$ refers to the order of $K_1$ as a differential operator.\\
We note that in all cases the determining equations (\ref{comutz1}) imply $f_m=k$ a constant and we can normalize $f_M=1$. We also note that in all cases we can add arbitrary powers of $H$ to the operator $K$. We shall omit case when $V(x)$ is constant ( e.g. $V_{a_1}$).\\
We are dealing with nonlinear ODEs of order $2 \leq n \leq 5$ which pass the Painlev\'e test \cite{Ablowitz:Pain}. These equations were analyzed in a series of articles by Chazy, Cosgrove et al. \cite{chazy11, cosgrove06, cosgrove00, Cosgrove:chazy}.
\newline
\newline
\textbf{I. Operator of type (a):}
\begin{align}\label{Va23}
&V_{a_2}=V,\\\nonumber
&K_{a_2}=p_x^2+\beta p_x+2V.\\
&V_{a_3}=\hbar ^2 \wp(x),\; f_2=\beta,\; f_1=3\hbar ^2\wp,\\\nonumber
&K_{a_3}=p_x^3+\beta p_x^2+3\hbar ^2\wp p_x+2\beta \hbar ^2 \wp-\frac{3}{2}i\hbar ^3 \wp'.
\end{align}
where $\wp(x)$ is the Weierstrass elliptic function. A special case for $V_{a_3}$ is
\begin{align}
V(x)=\frac{\hbar^{2}}{x^{2}},\;K_{a_{3}}=2p_{x}^{3}+\{\frac{3\hbar^{2}}{x^{2}},p_{x}\}.
\end{align}
\begin{align}\label{V45}
&V_{a_4}=\hbar ^2 \wp(x),\\\nonumber
&K_{a_4}=p_x^4+\beta p_x^3+4V p_x^2+(3\beta V-4 i \hbar V')p_x^2+(-\frac{3}{2}i \hbar \beta V'-8 V^2).
\end{align}
\begin{align}\label{V5}
&V_{a_5}=V,\\\nonumber
&K_{a_5}=p_x^5+\beta p_x^4+5V p_x^3+(-\frac{15}{2}i \hbar V'+4\beta V)p_x^2+(-\frac{25}{4} \hbar^2 V''-4 i \beta \hbar V'+\frac{15}{2}V^2)p_x\\\nonumber
&\quad \quad 
+\frac{15}{8} i \hbar^3 V^{(3)}-2 \beta\hbar ^2 V''-\frac{15}{2} i \hbar V(x) V'+4 \beta V^2.
\end{align}
In (\ref{V5}) $V$ satisfies the equation
\begin{equation}
\hbar^4 V^{(4)}-20 \hbar^2 V V''-10 \hbar^2 V'^2+40 V^3=0
\end{equation}
Setting $V=\hbar ^2U$, we get
\begin{equation}\label{a5}
U^{(4)}-20 U U''-10 U'^2+40 U^3=0
\end{equation}
This equation is a special autonomous case of the equation F-V, in \cite[p42]{cosgrove00}. It has the Painlev\'e property and it is solvable in terms of hyperelliptic functions.
The solution can be written as
\begin{align}\label{hypelli}
U=\frac{1}{4}(u_1+u_2)
\end{align}
where $u_1(x)$ and $u_2(x)$ are defined by inversion of the hyperelliptic integrals
\begin{align}\label{Va5}
&\int_{\infty}^{u_1(x)}{\frac{dt}{\sqrt{P(t)}}}+\int_{\infty}^{u_2(x)}{\frac{dt}{\sqrt{P(t)}}}=k_3,\\\nonumber
&\int_{\infty}^{u_1(x)}{\frac{t \; dt}{\sqrt{P(t)}}}+\int_{\infty}^{u_2(x)}{\frac{t\; dt}{\sqrt{P(t)}}}=x+k_4,
\end{align}
with $P(t)=t^5+32k_1t+k_2,$ where $k_1$ and $k_2$ are constants of integration.\\
The functions $u_1$ and $u_2$ are not meromorphic separately, each having movable quadratic branch points, however the solution $U$ is globally meromorphic. \\
\newline
\newline
\textbf{II. Operator of type (b):}
\begin{align}
&V_{b_{1}}=\frac{\alpha_1}{i\hbar} x,\\\nonumber
&K_{b_1}= p_x+\beta.\\
&V_{b_2}=-\frac{\alpha_1}{\beta \hbar}i x,\\\nonumber
&K_{b_2}= p_x^2+\beta p_x+2V.\\
&V_{b_3}=V,\; f_2=\beta ,\\\nonumber
&K_{b_3}=p_x^3+\beta_1 p_x^2+3V p_x+2\beta_1 V-\frac{3}{2}i\hbar V'.
\end{align}
where $V$ satisfies the first Painlev\'e equation
\begin{equation}
V''=\frac{6}{\hbar ^2}V^2+\frac{4i\alpha_1}{\hbar^3}x,
\end{equation}
and thus
\begin{equation}
V(x)=\hbar^{2}\omega_{1}^{2}P_{I}(\omega_{1}x),\; \omega_x=\frac{\sqrt[5]{4 i \alpha_1}}{\hbar}.
\end{equation}
\begin{align}
&V_{b_4}=V,\\\nonumber
&K_{b_4}=p_x^4+\beta p_x^3+4V p_x^2+3\beta V-4i\hbar V'-2\hbar ^2 V''-\frac{3}{2}i\hbar\beta V'+4V^2.
\end{align}
where $V$ satisfies the first Painlev\'e equation.
\begin{align}
&V_{b_5}=V,\\\nonumber
&K_{b_5}=p_x^5+\beta p_x^4+5V p_x^3+(-\frac{15}{2}i\hbar V'+4\beta V )p_x^2+(-\frac{25}{4} \hbar^2 V''-4 i \beta  \hbar V'+\frac{15}{2} V^2)p_x\\\nonumber
&\quad \quad 
+\frac{15}{8} i \hbar^3 V^{(3)}-2 \beta  \hbar^2 V''-\frac{15}{2} i \hbar V V'+4 \beta V^2.
\end{align}
The potential $V$ satisfies
\begin{equation}\label{Vb5}
\hbar^4 V^{(4)}-20 \hbar^2 V V''-10\hbar^2 V'^2+40V^3+\frac{16 i \alpha_x x}{\hbar}=0.
\end{equation}
Setting $V=\hbar ^2U,$ we get
\begin{equation}
U^{(4)}-20 U U''-10 U'^2+40 U^3+\frac{16 i \alpha_1 x}{h^7}=0
\end{equation}
This equation is also a special case of the equation F-V in \cite[p42]{cosgrove00}. The exact solution of it is not known and it is possible that its solution cannot be expressed in terms of classical transcendents nor one of the original Painlev\'e transcendents.
\newline
\newline
\textbf{III. Operator of type (c):}
\begin{align}
&V_{c_2}=\frac{\beta}{x^2},\\\nonumber
&K_{c_2}= p_x^2+\frac{\alpha_1}{2\hbar}i x p_x+\frac{2\beta_1}{x^2}.\\
&V_{c_3}=V,\\\nonumber
&K_{c_3}= p_x^3+\beta p_x^2+(3V+\frac{i}{2\hbar}x)p_x+2\beta V-\frac{3}{2}i\hbar V'.
\end{align}
Setting $V=\hbar^2 U-\frac{i\alpha_1}{2\hbar}x$, $U(x)$ is the solution of the following equation
$$U^{(3)}=12U U'-4\frac{i \alpha_1}{\hbar^3}x U'-\frac{2 \alpha_1 i}{\hbar^3},$$
It admits the first integral
\begin{equation}\label{c3}
2UU''-U'^2-8U^3+4\frac{ \alpha_1 i}{\hbar^3}x U^2=k
\end{equation}
where $k$ is an integration constant.
For $k=0,$ by the change of variables
$$x=\frac{1}{\lambda} X,\; U=\lambda^2 W^2;\; \lambda=\frac{i}{h}\sqrt[3]{\alpha_1},$$
we get a special case of the second Painlev\'e equation 
\begin{equation}
W''-2W^3-X W=0
\end{equation}
Therefore, the solution for $V(x)$ is
\begin{align*}
V(x)=-\alpha_1 ^{\frac{2}{3}}P_2^2-i\frac{\alpha_1}{2\hbar}x.
\end{align*}
with $P_2=P_2(\frac{i}{\hbar}\sqrt[3]{\alpha_1}x)$.\\
For $k\neq 0,$ by the following transformation
$$x=\lambda X , \;U=\sqrt{- k \lambda ^2}W;\;\lambda=\sqrt[3]{\frac{\hbar ^3}{2\alpha_x i}}$$
we transform (\ref{c3}) to
\begin{equation}
W''=\frac{W'^2}{2W}+4 \lambda ^2\sqrt{-k \lambda ^2} W^2-X W-\frac{1}{2W}
\end{equation}
which is Ince-XXXIV \cite[p340]{Ince:ode} with the solution
$$2 \lambda ^2 \sqrt{-k \lambda ^2}W=P_2'+P_2^2+\frac{1}{2}X$$
where $P_2$ satisfies the second Painlev\'e equation
$$P_2''=2P_2^3+XP_2-2 \lambda ^2 \sqrt{-k \lambda ^2}-\frac{1}{2}.$$
The solution for $V$ is
$$V(x)=\frac{(2\alpha_1 i)^\frac{2}{3}}{2}(P_2'+P_2^2),$$
for $P_2=P_2(\sqrt[3]{\frac{2\alpha_1 i}{\hbar ^3}}x)$.
\begin{align}
&V_{c_4}=V,\\\nonumber
&K_{c_4}=p_x^4+\beta p_x^3+4V p_x^2+(-4i\hbar V'+3\beta V +\frac{i \alpha_1}{2\hbar}x)p_x-2\hbar ^2V''-\frac{3}{2}i\beta \hbar V'+4V^2.
\end{align}
For $\beta=0,\; V=\frac{k}{x^2},$ and for $\beta \neq 0,$ setting $V=\hbar ^2 U-\frac{i\alpha_1}{6\beta \hbar}x, \; U(x)$ is the solution of the following equation
$$U^{(4)}=12UU''+12 U'^2+\frac{2i \alpha_1}{\beta \hbar^3}U'+\frac{2\alpha_1^2}{3\beta ^2\hbar^6}$$
which is again a special case of equation F-I \cite{cosgrove06}. Its solution can be expressed in terms of the second Painlev\'e transcendent.\\
\begin{align}
&V_{c_5}=V,\\\nonumber
&K_{c_5}=p_x^5+\beta p_x^4+5V p_x^3+(-\frac{15}{2}i\hbar V'+4\beta V )p_x^2+(-\frac{25}{4} \hbar^2 V''-4 i \beta  \hbar V'+\frac{15}{2} V^2+\frac{i\alpha_1}{2\hbar} x)p_x\\\nonumber
&\quad \quad 
+\frac{15}{8} i \hbar^3 V^{(3)}-2 \beta  \hbar^2 V''-\frac{15}{2} i \hbar V V'+4 \beta V^2.\nonumber
\end{align}
The potential $V$ satisfies
\begin{equation}
\hbar^5 V^{(5)}-20 \hbar^3 V V^{(3)}-40 \hbar ^3 V' V''+120 \hbar V^2 V'+8 i\alpha_1 x V'+16 i\alpha_1 V=0
\end{equation}
Setting $V=\hbar^6 U(X),\; X=\hbar ^2 x,$ we get
\begin{equation}\label{fif3}
U^{(5)}-20 U^{(3)} U+120 U^2 U'-40 U' U''+\frac{8 i\alpha_1 X }{\hbar^{15}}U'+\frac{16 i\alpha_1}{\hbar^{15}}U=0
\end{equation}
This equation is Fif-III in \cite[p25,eq 2.71]{cosgrove00} and it has the Painlev\'e property. A first integral of it is
\begin{align*}
2uu''-u'^2-8Uu^2+k=0
\end{align*}
where $u=U''-6U^2-\frac{2 i \alpha_1}{\hbar ^{15}}X.$\\
When $k=0,$ a particular solution of (\ref{fif3}) can be obtained by setting $u=0.$ This solution is $U=P_I,$ where $P_I(X)$ satisfies the Painlev\'e first equation.
\newline
\newline
\textbf{IV. Operator of type (d):}

\begin{align}
&V_{d_{1}}=\frac{\alpha_{1}^{2}}{2\hbar ^2} x^{2}, \\\nonumber
&K_{d_1}=p_x-\frac{\alpha_1}{\hbar}ix.\\
&V_{d_2}=\frac{{\alpha_{1}}^2}{8\hbar ^2}x^2+\frac{\beta}{x^2},\\\nonumber
&K_{d_2}=p_x^2-\frac{\alpha_1}{\hbar}i x p_x-\frac{\alpha_1^2}{\hbar ^2} x^2+\frac{2\beta}{x^2}.\\
&V_{d_3}=V(x),\\\nonumber
&K_{d_3}=p_x^3-\frac{\alpha_1}{\hbar}i x p_x^2+(3V-\frac{\alpha_1^2}{2\hbar ^2} x^2)p_x+(\frac{\hbar ^3}{4 \alpha }i V^{(3)}-\frac{3\hbar }{\alpha}i V V'-(\frac{5}{2} i \hbar -\frac{\alpha }{2\hbar } i x^2) V'+\frac{\alpha ^2}{2 \hbar} i  x).
\end{align}

Setting $V=\hbar ^2 U(x)+\frac{{\alpha_1}^2}{6 \hbar ^2}x^{2}-\frac{\alpha_1}{3}$, $U$ is the solution of
$$U^{(4)}=12UU''+12 U'^2-\frac{4 {\alpha_1}^2}{\hbar ^4}x U'-\frac{8 {\alpha_1}^2}{\hbar ^4}U-\frac{8{\alpha_1}^4}{3\hbar ^8}x^2$$
which is a special case of equation F-I \cite{cosgrove06}. The solution for $V(x)$ is

\begin{equation}\label{d3}
V(x)=\epsilon \alpha_1 P_4' +\frac{2\alpha_1 ^2}{\hbar ^2}( P_4^{2}+ x P_4)+\frac{\alpha_1 ^2 }{2\hbar ^2}x^{2}+(\epsilon - 1)\frac{\alpha_1}{3}-\frac{\hbar ^2}{6}k_{1},
\end{equation}

where $\epsilon=\pm 1$ and $P_4$ satisfies the fourth Painlev\'e equation

\begin{equation}
P_4''=\frac{(P_4')^{2}}{2P_4}+\frac{6\alpha_1^2}{\hbar ^4} P_4^{3} +\frac{8\alpha_1^2}{\hbar ^4} x P_4^{2}+(\frac{2\alpha_1^2}{\hbar ^4} x^{2}-k_{1})P_4 +\frac{k_{2}}{P_4}.
\end{equation}
$k_1$ and $k_2$ are integration constants.
\begin{align}
&V_{d_4}=V,\\\nonumber
&K_{d_4}=p_x^4-\frac{i \alpha x}{\hbar}p_x^3+(4V-\frac{\alpha^{2}x^{2}}{2 \hbar^{2}})p_x^2+f_1 p_x+f_0.
\end{align}
 
setting $u(x)=\int{V}dx,$ we get

\begin{align}
&f_{1}=-\frac{i x\alpha_{1}^{2}}{2\hbar} +\frac{i x^{3} \alpha_{1}^{3}}{6\hbar^{3}}-\frac{i \alpha_{1} u}{\hbar}-\frac{3 i x \alpha_{1} u'}{\hbar} -4 i \hbar u'', \\\nonumber
&f_{0}= -\frac{\alpha_{1}^{3} }{2\hbar^{2}} x^{2}+\frac{\alpha_{1}^{4}}{24\hbar^{4}} x^{4} -\frac{\alpha_{1}^{2}}{\hbar^{2}} x u+( \alpha_{x} -\frac{\alpha_{1}^{2} }{\hbar^{2}}x^{2})u' +4 u'^{2} -\frac{3}{2} \alpha_{1} x u'' -2 \hbar^{2} u'''.
\end{align}
thus $u(x)$ is the solution of the following equation

\begin{align}\label{d4}
0=& k -\alpha_{1}^{2}x^{2}+\frac{3 \alpha_{1}^{3}x^{4}}{4\hbar^{2}} -\frac{ \alpha_{1}^{4} x^{6}}{36 \hbar^{4}}+\frac{4 \alpha_{1}^{2} x^{3}}{3\hbar^{2}} u +2 u^{2} -2\hbar^{2} u' - 6\alpha_{1} x^{2} u'\\\nonumber
& +\frac{2\alpha_{1}^{2} x^{4}}{3\hbar^{2}} u' - 4  x u u' -  6  x^{2} u'^{2} + 2\hbar^{2} x u'' +\hbar^{2} x^{2} u'''.
\end{align}
By the following transformation
$$X=x^2, U=-\frac{x}{2 \hbar ^2}u+\frac{3 \hbar ^4-9 \alpha_{x}  h^2 x^2+\alpha_{x} ^2 x^4}{48 \hbar ^4}$$
we transform (\ref{d4}) to
\begin{align}\label{chazyI}
X^2U^{(3)}=-2(U'(3XU'-2U)-\frac{\alpha_{1}^2}{8\hbar ^4} X (X U'-U)+k_1X+k_2)-XU'',
\end{align}
where
$k_1=-\frac{7 \alpha_{1}^2}{256\hbar ^4} ,\;k_2=\frac{-4k-3\alpha_1\hbar ^2}{128 \hbar ^4}.$
The equation (\ref{chazyI}) is a special case of the Chazy class I equation \cite{chazy11, Cosgrove:chazy}. 
It admits the first integral
\begin{align}\label{SDIb}
X^2U''^2=-4(U'^2(XU'-U)-\frac{\alpha_{1}^2}{16\hbar ^4}(XU'-U)^2+k_1(XU'-U)+k_2U'+k_3)
\end{align}
where $k_3$ is the integration constant. The equation is the canonical form SD-I.b. 
The solution is
\begin{align}
U=&\frac{1}{4}(\frac{1}{P_{5}}(\frac{XP_5'}{P_{5}-1}-P_{5})^2-(1-\sqrt{2\lambda})^2(P_{5}-1)-2\beta \frac{P_{5}-1}{P_{5}}+\gamma X \frac{P_{5}+1}{P_{5}-1}+2 \delta \frac{X^2P_{5}}{(P_{5}-1)^2}),\nonumber\\
U'=&-\frac{X}{4P_{5}(P_{5}-1)}(P_{5}'-\sqrt{2\lambda} \frac{P_{5}(P_{5}-1)}{X})^2-\frac{\beta}{2X}\frac{P_{5}-1}{P_{5}}-\frac{1}{2}\delta X \frac{P_{5}}{P_{5}-1}-\frac{1}{4}\gamma,\nonumber\\
\end{align}
where $P_5=P_5(x^2),$ satisfies the fifth Painlev\'e equation
\begin{align}\label{P5}
P_5''=(\frac{1}{2P_5}+\frac{1}{P_5-1})P_5'^2-\frac{1}{X}P_5'+\frac{(P_5-1)^2}{X^2}(\lambda P_5+\frac{\beta}{P_5})+\gamma \frac{P_5}{X}+\delta \frac{P_5(P_5+1)}{P_5-1},
\end{align}
with
$$\alpha_{x}^2=-8 \hbar ^4\delta,\; k_1=-\frac{1}{4}(\frac{1}{4}\gamma ^2+2\beta \delta -\delta (1-\sqrt{2 \lambda})^2),\; k_2=-\frac{1}{4}(\beta \gamma+\frac{1}{2}\gamma (1-\sqrt{2\lambda})^2),$$
$$k_3=-\frac{1}{32}(\gamma ^2((1-\sqrt{2 \lambda})^2-2\beta)-\delta((1-\sqrt{2 \lambda})^2+2\beta)^2).$$
The solution for the potential is
\begin{align}
V(x)=&\frac{\alpha_1^2}{8\hbar ^2}x^2+\hbar ^2 \big( \frac{\gamma}{P_5-1}+\frac{1}{x^2}(P_5-1)(\sqrt{2\lambda }+\lambda(2P_5-1)+\frac{\beta}{P_5})\nonumber\\
&+x^2(\frac{P_5'^2}{2 P_5}-\frac{\alpha_1^2}{8\hbar ^4} P_5 )\frac{(2P_5-1)}{(P_5-1)^2}-\frac{P_5'}{P_5-1}-2\sqrt{2\lambda}P_5'\big)+\frac{3\hbar^2}{8x^2}.\nonumber\\
\end{align}
We could also choose the values of $k_i, i=1,2,3$ in a way to have (\ref{P5}) with the following parameter values:
$$\lambda=-\beta=\frac{B^2}{8}, \gamma=0, \delta=2A^2 \neq 0.$$
we can then reduce the fifth Painlev\'e equation with such parameters to a third Painlev\'e equation \cite[Thm 34.3(p.170),Thm 41.2(p.208), Thm 41.5(p.210)]{Gromak:pain}. Hence in this case we can obtain a solution in terms of third Painlev\'e transcendent.\\
\begin{align}
&V_{d_5}=V,\\\nonumber
&K_{d_5}=p_x^5-\frac{i\alpha_1}{\hbar}x p_x^4+(-\frac{\alpha_{1}^2 }{2\hbar^2}x^2 +5V)p_x^3+f_2 p_x^2+f_1p_x+f_0.
\end{align}
setting $u(x)=\int{V}dx,$ we get

\begin{align}
&f_2=-\frac{i \alpha_1 ^2}{2 \hbar}x +\frac{i \alpha_1 ^3}{6 \hbar^3}x^3-\frac{i \alpha_1}{\hbar}u-\frac{4 i\alpha_1}{\hbar}x u'-\frac{15}{2} i \hbar u'',\nonumber\\
&f_{1}=-\frac{\alpha_1 ^3 x^2}{2 \hbar^2}+\frac{\alpha_1 ^4}{24 \hbar^4}x^4-\frac{\alpha_1 ^2 }{\hbar ^2} x u+\alpha_1 u'-\frac{3\alpha_1 ^2}{2 \hbar^2} x^2 u'+\frac{15}{2} u'^2-4 \alpha_1  x u''-\frac{25}{4} \hbar ^2 u^{(3)}, \nonumber\\
&f_{0}=\frac{i}{48 \alpha_1  \hbar ^3} (-3 \hbar^8 u^{(6)}+114 \alpha_1  \hbar^6 u^{(4)}+60 \hbar^6 u^{(4)} u'+120 \hbar^6 u^{(3)} u''-6 \alpha_1 ^2 \hbar^4 x^2
   u^{(4)}+48 \alpha_1 ^2 \hbar^4 x u^{(3)}-96 \alpha_1 ^2 \hbar^4 u''\nonumber\\
   &-648 \alpha_1  \hbar^4 u' u''-360 \hbar^4 u'^2 u''+84\alpha_1 ^3 \hbar^2 x^2 u''+48 \alpha_1 ^2 \hbar^2 x u u''+96 \alpha_1 ^3 \hbar^2 x u'+72 \alpha_1 ^2 \hbar^2 x^2 u'u''+24
   \alpha_1 ^3 \hbar^2 u\nonumber\\
  & +36 \alpha_1 ^4 \hbar^2 x-2 \alpha_1 ^4 x^4 u''-4 \alpha_1 ^5 x^3).\nonumber\\
\end{align}
and
\begin{align}\label{d5}
&9 \hbar^{10}( x u^{(6)}- u^{(5)})+18 \hbar^6 x \left(-10 \hbar^2 u'+\alpha_1 ^2 x^2-4 \alpha_1 
   \hbar^2 \right)u^{(4)}\nonumber\\
   &+\hbar^6 \left(-360 \hbar^2 x u''+180 \hbar^2 u'+126 \alpha_1 ^2 x^2+72 \alpha_1 \hbar^2\right)u^{(3)}+90 \hbar^8 u''^2\nonumber\\
   &+\left(1080 \hbar^6 x u'^2+864 \alpha_1  \hbar^6 x u'-216 \alpha_1 ^2 \hbar^4 x^3 u'-144\alpha_1 ^2 \hbar^4 x^2 u+6 \alpha_1 ^4 \hbar^2 x^5-144 \alpha_1 ^3 \hbar^4 x^3+288 \alpha_1 ^2 \hbar^6 x\right)u''\nonumber\\
   &-360 \hbar^6 u'^3-432 \alpha_1  \hbar^6 u'^2-468 \alpha_1 ^2 \hbar^4 x^2 u'^2-144\alpha_1 ^2 \hbar^4 x u u'-288 \alpha_1 ^2 \hbar^6 u'-432 \alpha_1 ^3 \hbar^4 x^2 u'\nonumber\\
   &+42 \alpha_1 ^4 \hbar^2 x^4 u'+72 \alpha_1 ^2 \hbar^4 u^2+48 \alpha_1 ^4 \hbar^2 x^3 u-90 \alpha_1 ^4 \hbar^4 x^2+36\alpha_1 ^5 \hbar^2 x^4-\alpha_1 ^6 x^6=0\nonumber\\
\end{align}
This equation passes the Painlev\'e test. Substituting the Laurent series
$$u=\sum_{k=0}^{\infty} d_k(x-x_0)^{k+p}, \; d_0 \neq0,$$
in (\ref{d5}), we find $p=-1.$ The resonances are $r=1,2,5,6,8$ and we obtain $d_0=-\hbar ^2.$ The constants $d_1, d_2, d_5, d_6$ and $d_8$ are arbitrary, as they should be. It is not known whether (\ref{d5}) can be solved in terms of known functions.

\section{Classification of classical algebraic systems in one dimension}
We consider the Hamiltonian (\ref{H1}) and polynomial $K_1$ (\ref{Lx}) in classical mechanics and require that they satisfy one of the equations  (\ref{types}) where $[H_1,K_1] \equiv \{H_1,K_1\}_{PB}$ is now a Poisson bracket.\\
Instead of (\ref{comutDg}) we now have
\begin{align}
\{H_1,K_1\}_{PB}=\sum_{l=0}^{M+1}Z_l(x)p_{x}^{l}
\end{align}
with
\begin{align}
Z_0=f_1V',\;Z_l=(l+1)f_{l+1}V'-f'_{l-1}, 1 \leq l \leq M-1,\; Z_M=-f'_{M-1},\;Z_{M+1}=-f'_{M}.
\end{align}
For all the algebras in (\ref{types}) we obtain $f_{M}=k$ a constant, so we can set $f_M=1.$\\
\textbf{I. Polynomial of type (a):}\\
We have $Z_l=0, 0 \leq l \leq M$ and the result is trivial. In the case that $K_1$ is of order $1,3$ and $5$ we get a constant potential. For $K_1$ of order $2$ and $4$, the only function of $x$ and $p_{x}$ that Poisson commutes with the Hamiltonian is a function of $H_1$ itself. In particular the polynomial $K_1$ is any polynomial in $H.$ For all $M$ we find that either $V$ is constant or $K$ is a polynomial in $H.$\\
Notice that this is quite different from the quantum case (\ref{Va23}),...,(\ref{Va5}) where we obtain potentials expressed in terms of nonlinear special functions having the Painlev\'e property.\\
The types (b), (c) and (d) are more interesting and provide specific potentials that will generate superintegrable systems in $E_2.$\\
\textbf{II. Polynomial of type (b):}\\
The determining equations in this case are
\begin{align}
&f_1V'=\alpha_1, \;f'_{M-1}=0,\\\nonumber
&(l+1)f_{l+1}V'-f'_{l-1}=0, 1 \leq l \leq M-1,
\end{align} 
They were already solved by G\"{u}ng\"{o}r et al \cite{Gungor:14}. For completeness we reproduce some of their results in our notations.\\
\begin{align}
&V_{b_{1}}=\alpha_1 x,\\\nonumber
&K_{b_1}=p_x+\beta.\\
&V_{b_{2}}=\frac{\alpha_1}{\beta}x,\\\nonumber
&K_{b_2}=2H_1+\beta p_x.\\
&V_{b_{3}}=\epsilon \sqrt{(\frac{2\alpha_1}{3}x)}; \; \epsilon=\pm 1; \\\nonumber
&K_{b_3}=p_x(2H_1+V)+2\beta H_1.\\
&V_{b_{4}}=\epsilon \sqrt {(\frac{2\alpha_1}{3\beta}x)};\; \epsilon =\pm 1;\\\nonumber
&K_{b_4}=4H_1^2+2\beta p_x H_1+\beta V p_x.\\
&V_{b_{5}}=\sqrt[3]{(\frac{2\alpha_1}{5}x)},\\\nonumber
&K_{b_5}=4p_x H_1^2+4\beta h_1^2+2Vp_x H_1+\frac{3}{2}V_2 p_x.
\end{align}
\textbf{III. Polynomial of type (c):}\\
The determining equations are
\begin{align}
&f_1V'=\alpha_1 V,\;3f_{3}V'-f'_{1}=\frac{\alpha_1}{2},\;f'_{M-1}=0, M \neq 2,\\\nonumber 
&(l+1)f_{l+1}V'-f'_{l-1}=0, 1 \leq l \leq M-1, l\neq 2. 
\end{align}
The case when $K_1$ is a first order polynomial does not exist. The solutions for $2 \leq M \leq 5$ are
\begin{align}
&V_{c_{2}}=\frac{c}{x^{2}},\\\nonumber
&K_{c_2}= 2H_1-\frac{\alpha_1}{2} x p_x.\\
&V_{c_{3}}=V,\;  (\alpha_1 x-2V)^{2}V=c,\\\nonumber
&K_{c_3}= p_x^3+\beta p_x^2+(3V-\frac{\alpha_1}{2}x)p_x+2\beta V.\\
&V_{c_{4}}=V;\; (2\beta V- \alpha_1 x)^{2}V=c,\\\nonumber
&K_{c_4}=p_x^4+\beta p_x^3+4V p_x^2+(3\beta V-\frac{\alpha_1}{2}x)p_x+4 V^2.\\
&V_{c_{5}}=V;\; -(\alpha_1 x-15 V^2) V'=2\alpha_1 V,\\\nonumber
&K_{c_5}=p_x^5+\beta p_x^4+5V p_x^3+4 \beta V p_x^2+(\frac{15}{2} V^2-\frac{\alpha_1}{2}x)p_x+4 \beta V^2.
\nonumber
\end{align}

\textbf{IV. Polynomial of type (d):}\\
In this case we define polynomial ladder operators as
\begin{align}
K_1^{\pm}=\sum_{l=0} ^M f_l p_x^l
\end{align}
where $f_l= c g_l$, with
\begin{align}
c =
\begin{cases}
\mp i \quad \text{for}\; l \; \text{even} \\
 1 \quad \text{for}\; l \;\text{odd}
 \end{cases}
\end{align} 
 and they satisfy the algebraic relations
$$\{H_1,K_1^{\pm} \}=\pm i \alpha_1 K_1 ^{\pm}$$
The determining equations are
\begin{align}
&g_1V'=\alpha_1 g_0,\;g'_{M-1}=(-1)^{M-1}\alpha_1 g_M,\\\nonumber
&(l+1)g_{l+1}V'-g'_{l-1}=-\alpha_1 g_l, 1 \leq l \leq M-1. 
\end{align}
Their solutions are
\begin{align}
&V_{d_{1}}=\frac{\alpha_{1}^{2}x^{2}}{2},\\\nonumber
&K_{d_1}=p_x+\alpha_1 x.\\
&V_{d_{2}}=\frac{\alpha_{1}x^{2}}{8}+\frac{\gamma}{x^{2}},\\\nonumber
&K_{d_2}=p_x^2-\alpha_1 x p_x+2V-\frac{\alpha_1^2}{2} x^2.\\
&V_{d_{3}}=V,\\\nonumber
&K_{d_3}=p_x^3+\alpha_1 x p_x^2+(3V-\frac{\alpha_1^2}{2} x^2)p_x-(\frac{(\alpha_1^2 x^2-6V)}{2 \alpha_1}V'.\nonumber
\end{align}
The potential $V$ satisfies
\begin{align}\label{d3-2}
24xVV'-4 \alpha_1 ^2 x^3 V'-12 V^2-12 \alpha_1 ^2 x^2 V+\alpha_1 ^4 x^4+4d=0.
\end{align}
It admits the following first integral
\begin{align}\label{d3-1}
9V^{4}-14\alpha_{1}^{2} x^{2} V^{3}+(\frac{15}{2}\alpha_{1}^{4}x^{4}- 6d)V^{2}
-2\alpha_1^2 x^2(\frac{3}{4}\alpha_{1}^{4}x^{4}-d)V +(\frac{\alpha_1^8}{16}x^8+\frac{1}{2}d \alpha_{1}^{4}x^{4}+d^{2})=0.
\end{align}
\newline
\begin{align}
&V_{d_{4}}=V,\\\nonumber
&K_{d_4}=p_x^4-\alpha_1 x p_x^3+(4u'-\frac{\alpha_1^2}{2} x^2)p_x^2+(\frac{\alpha_1 ^3}{6}x^3-\alpha_1 u-3\alpha_1 x u')p_x+(\frac{\alpha_1 ^2}{6}x^3-u-3xu')u'',
\end{align}
where $u(x)=\int{V}dx$ and $u$ satisfies
\begin{align}\label{d4-2}
&3 x^2 u'^2+2 x u u'-\frac{1}{3} \alpha_1 ^2 x^4 u'-u^2-\frac{2}{3} \alpha_1 ^2 x^3 u+\frac{\alpha_1^4 x^6}{72}+k_1x+k_2=0.
\end{align}
Also using (\ref{KKd}) we find that $V(x)$ satisfies the following $5th$ order algebraic equation
\begin{align}\label{d4-1}
&0=-32 \alpha_{1}^{2}x^{2}V^{5}+(128d+17 \alpha_{1}^{4}x^{4})V^{4}\\
&+(-96c-\frac{128d^{2}}{\alpha_{1}^{2}x^{2}}-\frac{512e}{\alpha_{1}^{2}x^{2}}-32d\alpha_{1}^{2}x^{2}-\frac{7\alpha_{1}^{6}x^{6}}{2})V^{3} \nonumber \\  
&+(-40d^{2}+352e+\frac{256c^{2}}{\alpha_{1}^{4}x^{4}}+\frac{256cd}{\alpha_{1}^{2}x^{2}}+32c\alpha_{1}^{2}x^{2}+3d\alpha_{1}^{4}x^{4}+\frac{11\alpha_{1}^{8}x^{8}}{32})V^{2}\nonumber \\
&+(-16cd-\frac{128cd^{2}}{\alpha_{1}^{4}x^{4}}-\frac{512ce}{\alpha_{1}^{4}x^{4}}-\frac{256c^{2}}{\alpha_{1}^{2}x^{2}}+\frac{64d^{3}}{\alpha_{1}^{2}x^{2}}+\frac{256de}{\alpha_{1}^{2}x^{2}}-\frac{\alpha_{1}^{10}x^{10}}{64})V \nonumber \\
&+64c^{2}+4d^{3}-112de+\frac{16d^{4}}{\alpha_{1}^{4}x^{4}}+\frac{128d^{2}e}{\alpha_{1}^{4}x^{4}}+\frac{256e^{2}}{\alpha_{1}^{4}x^{4}}-\frac{64cd^{2}}{\alpha_{1}^{2}x^{2}}+\frac{256ce}{\alpha_{1}^{2}x^{2}} \nonumber \\
&-8cd\alpha_{1}^{2}x^{2}+\frac{3}{2}d^{2}\alpha_{1}^{4}x^{4}+\frac{17}{2}e\alpha_{1}^{4}x^{4}-\frac{1}{4}c\alpha_{1}^{6}x^{6}+\frac{1}{64}d\alpha_{1}^{8}x^{8}+\frac{\alpha_{1}^{12}x^{12}}{4096}.\nonumber\\
&f_{3}=-\alpha_{1}x,\; f_{2}=(-\frac{1}{2}\alpha_{1}^{2}x^{2}+4V),\; f_{1}=\frac{1}{8\alpha_{1}x}(16d +\alpha_{1}^{4}x^{4}+8f_{0}-16\alpha_{1}^{2}x^{2}V-32V^{2}),\nonumber\\
&f_{0}=(\sqrt{-16 e +16 c V -16 d V^{2}+16 V^{4}}).\nonumber
\end{align}

\begin{align}
&V_{d_{5}}=V,\\\nonumber
&K_{d_5}=p_x^5+\alpha_1 x p_x^4+(5u'-\frac{\alpha_1^2}{2} x^2)p_x^3+(4 \alpha_1  x u'+\alpha_1 u-\frac{1}{6}\alpha_1 ^3 x^3)p_x^2\\\nonumber
&\quad \quad +\frac{1}{24}(180 u'^2-36\alpha_1 ^2 x^2 u'-24 \alpha_1^2 x u+\alpha_1 ^4 x^4)p_x\\\nonumber
&\quad \quad +\frac{1}{24 \alpha_1}(\alpha_1 ^4 x^4-24\alpha_1 ^2 x u-36\alpha_1 ^2 x^2 u'+180 u'^2) u''
\end{align}

where $u(x)=\int{V}dx$ and $u$ satisfies
\begin{align}\label{d5-1}
&(\alpha_1 ^4 x^4-24 \alpha_1 ^2 x u-36 \alpha_1 ^2 x^2 u'+180 u'^2)x u''-60u'^3-78\alpha_{1}^2 x^2 u'^2-24 \alpha_{1}^2 x u u'+7\alpha_{1}^4 x^4 u'\\
&+12 \alpha_{1}^2 u^2+8 \alpha_{1}^4 x^3 u-\frac{1}{6}\alpha_{1} ^6 x^6=0\nonumber
\end{align}
We could also use (\ref{KKd}) to get an algebraic equation for $V$, but it is not very illuminating.\\

The solutions for $H_{d_{3}}$ and $H_{d_{4}}$ are presented in \cite{marquette10b,marquette12}.\\
The list of Hamiltonians reduces to $H_{b_{1}}\equiv H_{b_{2}}$, $H_{d_{1}}$, $H_{c_{2}}$, $H_{d_{2}}$, $H_{a_{3}}\equiv H_{a_{4}}$, $H_{b_{3}}\equiv H_{b_{4}}$, $H_{c_{3}}\equiv H_{c_{4}}$, $H_{d_{3}}$, $H_{d_{4}}$, $H_{a_{5}}$, $H_{b_{5}}$, $H_{c_{5}}$, $H_{d_{5}}.$\\

\section{ Classification of superintegrable systems up to fifth order integrals : quantum and classical systems}

In Section 3 we classified all quantum algebraic systems in one dimension with $M$ satisfying $1 \leq M \leq 5$. Here we shall use $2$ copies of these algebras $H_1, K_1$ and $H_2,K_2$ to construct two dimensional superintegrable systems as described in Section 2 and Table $1$.\\
In Table $3$, column $1$ gives the order of the operator $K$ in the two dimensional systems. In column $2$ all entries have the form $(z_i,w_j)$. The letters $z$ and $w$ run through the type $a,b,c,d$ as in (\ref{types}) with $z$ referring to the $x$ variable and $w$ to the $y$. The indices $i$ and $j$ correspond to the orders of the corresponding operators $K_1$ and $K_2$ and run between $1$ and $5.$ The values $(\alpha_1,\alpha_2)$ in column $3$ refer to the value of $\alpha_1$ in (\ref{typesb}), (\ref{typesc}) and (\ref{typesd}) in $x$-space and $y$-space, respectively.

\begin{table}
[h!]
\begin{tabular}{|l|l|l|}
  \hline
order of  &  type  &  $(\alpha_{1},\alpha_{2})$   \\[0.05cm]
integrals &    &    \\[0.05cm]
  \hline
	 1  & $(b_{1},b_{1})$ & $(\alpha_{1},\alpha_{2})$   \\[0.05cm]
\cline{2-3}	
        & $(d_{1},d_{1})$  & $(\alpha,\alpha)$  \\[0.05cm]
 \hline
 2  & $(d_{1},d_{1})$  & $(\alpha,2\alpha)$   \\[0.05cm]
 \cline{2-3}
    & $(d_{1},d_{2})$  & $(\alpha,\alpha)$   \\[0.05cm]
 \hline 
 3  & $(a_{3},a_{3})$ & $(\hbar^{2}, \hbar^{2})$  \\[0.05cm]
 \cline{2-3}	
   & $(b_{3},b_{3})$ & $(\alpha_{1},\alpha_{2})$  \\[0.05cm]
   \cline{2-3}	
   & $(b_{1},b_{3})$ & $(\alpha_{1},\alpha_{2})$   \\[0.05cm]
   \cline{2-3}	
  & $(c_{2},b_{1})$ & $(\alpha_{1},\alpha_{2})$  \\[0.05cm]
  \cline{2-3}	
   & $(c_3,b_1)$ & $(\alpha_{1},\alpha_{2})$   \\[0.05cm]
   \cline{2-3}	
   & $(d_{1},d_{3})$  & $(\alpha,\alpha)$  \\[0.05cm]
  \cline{2-3}		
 & $(d_{1},d_{2})$  & $(\alpha,2\alpha)$  \\[0.05cm]
   \cline{2-3}		
  & $(d_{2},d_{2})$  & $(\alpha,\alpha)$  \\[0.05cm]
   \cline{2-3}
   & $(d_{1},d_{1})$  & $(\alpha,3\alpha)$  \\[0.05cm]
     \hline
4  & $(d_{4},d_{1})$ & $(\alpha,\alpha)$   \\[0.05cm]
  \cline{2-3}
  & $(d_{3},d_{2})$ & $(\alpha,\alpha)$    \\[0.05cm]
   \cline{2-3}
  & $(d_{1},d_{3})$ &  $(\alpha,2\alpha)$    \\[0.05cm]
   \cline{2-3}
  & $(d_{1},d_{2})$ &  $(\alpha,3\alpha)\&(2\alpha,\alpha)$   \\[0.05cm]
  \cline{2-3}
   & $(d_{1},d_{1})$ & $(\alpha,4\alpha)\&(2\alpha,3\alpha)$   \\[0.05cm]
  \hline
  					
\end{tabular}
\quad
\begin{tabular}{|l|l|l|}
  \hline
order of  &  type  &  $(\alpha_{1},\alpha_{2})$   \\[0.05cm]
integrals &    &    \\[0.05cm]
  \hline
   5  & $(a_{3},a_{5})$ & $....$  \\[0.05cm]
 \cline{2-3}	
   & $(a_{5},a_{5})$ & $....$  \\[0.05cm]
   \cline{2-3}	
   & $(b_{1},b_{5})$ & $(\alpha_{1},\alpha_{2})$   \\[0.05cm]
   \cline{2-3}	
   & $(b_{3},b_{5})$ & $(\alpha_{1},\alpha_{2})$   \\[0.05cm]
   \cline{2-3}	
   & $(b_{5},b_{5})$ & $(\alpha_{1},\alpha_{2})$   \\[0.05cm]
   \cline{2-3}	
   & $(c_{3},c_{3})$ & $(\alpha_{1},\alpha_{2})$   \\[0.05cm]
   \cline{2-3}	
   & $(c_{2},c_{3})$ & $(\alpha_{1},\alpha_{2})$   \\[0.05cm]
    \cline{2-3}	
   & $(c_{2},b_{3})$ & $(\alpha_{1},\alpha_{2})$   \\[0.05cm]
    \cline{2-3}	
   & $(c_{3},b_{3})$ & $(\alpha_{1},\alpha_{2})$   \\[0.05cm]
    \cline{2-3}	
   & $(c_{5},b_{1})$ & $(\alpha_{1},\alpha_{2})$   \\[0.05cm]
   \cline{2-3}	
   & $(c_{5},b_{3})$ & $(\alpha_{1},\alpha_{2})$   \\[0.05cm]	
   \cline{2-3}
 & $(d_{5},d_{1})$ & $(\alpha,\alpha)$   \\[0.05cm]
  \cline{2-3}
   & $(d_{4},d_{2})$ & $(\alpha,\alpha)$   \\[0.05cm]
    \cline{2-3}
    & $(d_{3},d_{3})$ & $(\alpha,\alpha)$    \\[0.05cm]
   \cline{2-3}
  & $(d_{2},d_{2})$ &  $(\alpha,2\alpha)$   \\[0.05cm]
  \cline{2-3}
  & $(d_{1},d_{4})$ & $(\alpha,2\alpha)$   \\[0.05cm]
  \cline{2-3}
  & $(d_{1},d_{3})$ &  $(\alpha,3\alpha)$    \\[0.05cm]
  \cline{2-3}
  & $(d_{1},d_{2})$ &  $(\alpha,4\alpha)\& (\alpha,\alpha)$   \\[0.05cm]
  \cline{2-3}
   & $(d_{1},d_{1})$ & $(\alpha,5\alpha)\&(\alpha,2\alpha)\&(\alpha,\alpha)$   \\[0.05cm]
    \hline	
\end{tabular}
\caption{\footnotesize{Classification of superintegrable systems}}
\end{table}
\newpage
The systems in $E_2$ can however admit lower order integrals of motion as this construction does not necessarily provide integrals of the lowest order. As an example for one of the Smorodinsky-Winternitz potential the ladder operators lead to an integral of order $3$ that is in fact the commutator or Poisson commutator of two integrals of order $2$. The same phenomena occurs for some of the Gravel potentials.\\
The potential $(Q.10)$ in Gravel's list can be obtained by $(d_2,d_2)$ construction of $8$th order integral with $(\alpha_1,\alpha_2)=(\alpha, 3\alpha).$ The potentials $Q_1^1, Q_3^2, Q_3^3$ and $Q_3^4$ in \cite{sajedi17} are the cases $(d_4,d_2),(c_2,c_3),(c_2,b_3)$ and $(a_3,a_5)$ respectively.\\
Let us present the list of obtained quantum superintegrable systems.\\
\textbf{Quantum superintegrable systems that occur in infinite families:}\\
\textbf{Jauch-Hill potentials:}\\
These anisotropic harmonic oscillator potentials have the form $V=\omega ^2 (n x^2+m y^2)$ where $n$ and $m$ are two mutually prime positive integers \cite{hill40}.\\
$(d_1,d_1):$\\
\begin{align*}
&V=\frac{\alpha^{2}}{2\hbar ^2} (x^{2}+4 y^{2}),\\
&K=(K_1^{\dagger})^2(K_2^{-})-(K_1^{-})^2(K_2^{\dagger})=(y p_x-x p_y)p_x.
\end{align*}
\begin{align*}
&V=\frac{\alpha^{2}}{2\hbar ^2} (x^{2}+9y^2)\\
&K=(K_1^\dagger)^3(K_2^{-})-(K_1^{-})^3(K_2^\dagger)=(xp_y-y p_x)p_x^2.
\end{align*}
\begin{align*}
V=\frac{1}{2\hbar ^2} (\alpha_x^{2}x^{2}+\alpha_y^{2} y^2),
\end{align*}
\begin{align*}
&(\alpha_1,\alpha_2)=(\alpha,4\alpha):\;K=(K_1^\dagger)^4(K_2^{-})-(K_1^{-})^4(K_2^\dagger)=(xp_y-y p_x)p_x^3.\\
&(\alpha_1,\alpha_2)=(2\alpha,3\alpha):\;K=(K_1^\dagger)^3(K_2^{-})^2-(K_1^{-})^3(K_2^\dagger)^2=(x p_y-yp_x)p_x^2p_y.
\end{align*}
\begin{align*}
&(\alpha_1,\alpha_2)=(\alpha,\alpha):\;K=(K_1^\dagger)^3(K_2^{-})^3-(K_1^{-})^3(K_2^\dagger)^3=(x p_y-y p_x)p_x^2 p_y^2.\\
&(\alpha_1,\alpha_2)=(\alpha,2\alpha):\;K=(K_1^\dagger)^4(K_2^{-})^2-(K_1^{-})^4(K_2^\dagger)^2=(x p_y-y p_x)p_x^3 p_y.\\
&(\alpha_1,\alpha_2)=(\alpha,5\alpha):\;K=(K_1^\dagger)^5(K_2^{-})-(K_1^{-})^5(K_2^\dagger)=(x p_y-y p_x)p_x^4.
\end{align*}
All values of $n$ and $m$ can be obtained in this manner. The $\hbar$ in the denominator of $V$ has no meaning since we can have $\alpha_1^2=n \hbar^2,\; \alpha_2^2=m \hbar^2.$\\ 
\textbf{Smorodinsky-Winternitz potentials:}\\
The original multiseparable potentials in $E_2$ were
\begin{align}\label{SW1}
V(x,y)=\omega ^2(x^2+y^2)+\frac{\beta}{x^2}+\frac{\gamma}{y^2}
\end{align}
that is separable in Cartesian, polar and elliptic coordinates, and
\begin{align}\label{SW2}
V(x,y)=\omega ^2(x^2+4y^2)+\frac{\gamma}{y^2}
\end{align} 
that is separable in Cartesian and parabolic coordinates.\\
Both allow second order integrals of motion \cite{Fris:Fri, Makanov:Nonrel}. These, plus two further ones, not allowing separation in Cartesian coordinates, were later called Smorodinsky-Winternitz potentials \cite{Evans:CM, Evans:W, Evans:SW, Gros95}. For $\omega =0$ "degenerate" forms of the Smorodinsky-Winternitz potentials exist, such as
\begin{align}\label{SW3}
V(x,y)=\alpha y+\frac{\beta}{x^2}+\frac{\gamma}{y^2}.
\end{align}
In the present approach the potentials (\ref{SW1}) and (\ref{SW2}) are built into infinite sets of potentials generalizing the Smorodinsky-Winternitz potentials both in classical and quantum mechanics \cite{RTW08, EV08}. They occur when we consider the cases $(d_1,d_2)$ and $(d_2,d_2).$
\begin{align}\label{SW4}
&V=\omega ^2 (n^2 x^{2}+m^2 y^2)+\frac{\beta}{x^2}+\frac{\gamma}{y^2},\\
&K=(K_1^\dagger)^m(K_2^{-})^n-(K_1^{-})^m(K_2^\dagger)^n
\end{align}
Taking $m=n \geq 1, \; \omega \neq 0,\; \beta \neq 0$ and $\gamma \neq 0,$ we obtain the potential (\ref{SW1}). Taking $m=2n, \beta=0$ we obtain (\ref{SW2}). The pair $(d_1,d_2)$ provides (\ref{SW4}) with $\beta=0.$ The degenerate one (\ref{SW3}) is obtained as $(c_2,b_1).$ The potentials (\ref{SW4}) have been called "caged harmonic oscillators" \cite{EV08}.\\ 
\textbf{Elliptic and hyperelliptic function potentials:}\\
$(a_3,a_3):$
\begin{align*}
&V=\hbar ^2 (\wp(x)+\wp(y)),\\
&K=K_1+K_2.
\end{align*}
$(a_3,a_5)$:\\
\begin{align*}
V=\hbar ^2 \wp(x)+\hbar ^2 g(y),\;K=K_1+K_2
\end{align*}
$(a_5,a_5)$:\\
\begin{align*}
V=\hbar ^2 (f(x)+g(y)),\;K=K_1+K_2
\end{align*}
where $f(x)$ and $g(y)$ are hyperelliptic functions  satisfying equation (\ref{a5}) and are defined in (\ref{hypelli}).\\
\textbf{Potentials in terms of the first Painlev\'e transcendent:}\\
$(b_3,b_3):$
\begin{align*}
&V=\hbar^{2}(\omega_{1}^{2}P_{I}+\omega_{2}^{2}P_{I}),\\
&K=\alpha_2 K_1-\alpha_1 K_2.
\end{align*}
$(b_1,b_3):$
\begin{align*}
&V=\frac{\alpha_1}{i \hbar }x+\hbar^{2}(\omega_{2}^{2}P_{I}),\\
&K=\alpha_2 K_1-\alpha_1 K_2.
\end{align*}
$(c_2,b_3)$:\\
\begin{align*}
V=\frac{\beta}{x^2}+\hbar^{2}\omega_{2}^{2}P_{I}(\omega_{2}y),\; \omega_2=\frac{\sqrt[5]{4 i \alpha_2}}{\hbar},\;
K=\alpha_2 K_1-\alpha_1 H_1 K_2.
\end{align*}
\textbf{Potentials in terms of the second Painlev\'e transcendent:}\\
$(c_3,b_1):$
\begin{align*}
&V=-\alpha_1 ^{\frac{2}{3}}P_2^2-i\frac{\alpha_1}{2\hbar}x+\frac{\alpha_1}{i\hbar} y,\;P_2=P_2(\frac{i}{\hbar}\sqrt[3]{\alpha_1}x)\\
\text{and}\\
&V=\frac{\hbar^{2}}{2} (\epsilon  P_{2}'+ P_{2}^{2})+\frac{\alpha_2}{i\hbar} y,\;P_2=P_2(\sqrt[3]{\frac{2\alpha_1 i}{\hbar ^3}}x),
\end{align*}
\begin{align*}
K=\alpha_2 K_1-\alpha_1 H_1 K_2=\alpha_2 p_x^3-\frac{\alpha_1}{2} p_x^2 p_y.
\end{align*}
$(c_3,c_3)$:\\
\begin{align*}
&V=\hbar ^2 (f(x)+g(y))-\frac{i}{2 \hbar}(\alpha_1 x+\alpha_2 y),\; K=\alpha_2 H_2 K_1-\alpha_1 H_1 K_2=(\alpha_2 p_x -\alpha _1 p_y)p_x^2p_y^2.
\end{align*}
$(c_2,c_3)$:\\
\begin{align*}
&V=\frac{\beta}{x^2}+\hbar ^2 g(y)-\frac{\alpha_2 i}{2 \hbar}  y,\;K=\alpha_2 H_2 K_1-\alpha_1 H_1 K_2.
\end{align*}
$(c_3,b_3)$:\\
\begin{align*}
V=f(x)+\hbar^{2}\omega_{2}^{2}P_{I}(\omega_{2}y),\; \omega_y=\frac{\sqrt[5]{4 i \alpha_2}}{\hbar},\;
K=\alpha_2 K_1-\alpha_1 H_1 K_2
\end{align*}
where $f(x)$ and $g(y)$ satisfy equation (\ref{c3}).\\
\textbf{Potentials in terms of the fourth Painlev\'e transcendent:}\\
$(d_1,d_3):$
\begin{align*}
&V=\frac{\alpha^{2}}{2\hbar ^2} (x^{2}+y^{2})+\epsilon \alpha P_4' +\frac{2\alpha ^2}{\hbar ^2}( P_4^{2}+ y P_4),\; K=(K_1^\dagger)(K_2^{-})-(K_1^{-})(K_2^\dagger).
\end{align*}
$(d_3,d_2):$
\begin{align*}
&V=\frac{\alpha ^2 }{8\hbar ^2}(4x^{2}+y^2)+\frac{\beta}{y^2}+\epsilon \alpha P_4' +\frac{2\alpha ^2}{\hbar ^2}( P_4^{2}+ x P_4),\\
&K=(K_1^\dagger)(K_2^{-})-(K_1^{-})(K_2^\dagger)=(x p_y-y p_x)p_x^2 p_y.
\end{align*}
$(d_1,d_3):$
\begin{align*}
&V=\frac{\alpha^{2}}{2\hbar ^2} (x^{2}+4y^{2})+2 \epsilon \alpha P_4' +\frac{8\alpha ^2}{\hbar ^2}( P_4^{2}+ y P_4),\; K=(K_1^\dagger)^2(K_2^{-})-(K_1^{-})^2(K_2^\dagger)=(x p_y-y p_x)p_x p_y^2.
\end{align*}
\begin{align*}
V=\frac{\alpha^{2}}{2\hbar ^2}( x^{2}+9y^{2})+3 \epsilon \alpha P_4' +\frac{18\alpha ^2}{\hbar ^2}( P_4^{2}+ y P_4),\;K=(K_1^\dagger)^3(K_2^{-})-(K_1^{-})^3(K_2^\dagger)=(x p_y -y p_x)p_x^2p_y^2
\end{align*}
$(d_3,d_3)$:\\
\begin{align*}
V=f(x)+g(y),\;K=(K_1^\dagger)(K_2^{-})-(K_1^{-})(K_2^\dagger)=(x p_y-y p_x)p_x^2p_y^2
\end{align*}
where $f(x)$ and $g(y)$ are given in (\ref{d3}) and $\alpha_1=\alpha_2=\alpha$.\\
\textbf{Potentials in terms of the fifth Painlev\'e transcendent:}\\
$(d_4,d_1):$
\begin{align*}
&V=\frac{\alpha ^2}{8\hbar ^2}(x^2+4 y^{2})+\frac{3\hbar^2}{8x^2}+\hbar ^2 \big( \frac{\gamma}{P_5-1}+\frac{1}{x^2}(P_5-1)(\sqrt{2\lambda }+\lambda(2P_5-1)+\frac{\beta}{P_5})\nonumber\\
&\quad +x^2(\frac{P_5'^2}{2 P_5}-\frac{\alpha ^2}{8\hbar ^4} P_5 )\frac{(2P_5-1)}{(P_5-1)^2}-\frac{P_5'}{P_5-1}-2\sqrt{2\lambda}P_5'\big),\\
&K=(K_1^\dagger)(K_2^{-})-(K_1^{-})(K_2^\dagger)=(x p_y-y p_x) p_x^3.
\end{align*}
$(d_4,d_2)$:\\
\begin{align*}
V=&\frac{\alpha^2}{8\hbar ^2}(x^2+y^2)+\frac{3\hbar^2}{8x^2}+\frac{\beta}{y^2}+\hbar ^2 \big( \frac{\gamma}{P_5-1}+\frac{1}{x^2}(P_5-1)(\sqrt{2\lambda }+\lambda(2P_5-1)+\frac{\beta}{P_5})\nonumber\\
&\quad +x^2(\frac{P_5'^2}{2 P_5}-\frac{\alpha^2}{8\hbar ^4} P_5 )\frac{(2P_5-1)}{(P_5-1)^2}-\frac{P_5'}{P_5-1}-2\sqrt{2\lambda}P_5'\big),\\
K=&(K_1^\dagger)(K_2^{-})-(K_1^{-})(K_2^\dagger)=(x p_y-y p_x) p_x^3 p_y.
\end{align*}
$(d_1,d_4)$:\\
\begin{align*}
V=&\frac{\alpha^{2}}{2\hbar ^2}( x^{2}+y^2)+\frac{3\hbar^2}{8y^2}+\hbar ^2 \big( \frac{\gamma}{P_5-1}+\frac{1}{y^2}(P_5-1)(\sqrt{2\lambda }+\lambda(2P_5-1)+\frac{\beta}{P_5})\nonumber\\
&\quad +y^2(\frac{P_5'^2}{2 P_5}-\frac{\alpha^2}{2\hbar ^4} P_5 )\frac{(2P_5-1)}{(P_5-1)^2}-\frac{P_5'}{P_5-1}-2\sqrt{2\lambda}P_5'\big)\\
K=&(K_1^\dagger)^2(K_2^{-})-(K_1^{-})^2(K_2^\dagger)=(x p_y-y p_x)p_x p_y^3.
\end{align*}
\textbf{Potentials satisfying higher order nonlinear equations passing the Painlev\'e test:}\\
$(b_1,b_5)$:\\
\begin{align*}
V=\frac{\alpha_1}{i\hbar} x+g(y),\; K=\alpha_2 K_1-\alpha_1 K_2,
\end{align*}
$(b_3,b_5)$:\\
\begin{align*}
&V=\hbar^{2}\omega_{1}^{2}P_{I}(\omega_{1}x)+g(y),\; \omega_1=\frac{\sqrt[5]{4 i \alpha_1}}{\hbar}.\\
&K=\alpha_2 K_1-\alpha_1 K_2.
\end{align*}
$(b_5,b_5)$:\\
\begin{align*}
&V=f(x)+g(y),\; K=\alpha_2 K_1-\alpha_1 K_2.
\end{align*}
where $f(x)$ and $g(y)$ satisfy equation (\ref{Vb5}).\\
$(d_5,d_1)$:\\
\begin{align*}
&V=f(x)+\frac{\alpha^{2}}{2\hbar ^2} y^{2},\;K=(K_1^\dagger)(K_2^{-})-(K_1^{-})(K_2^\dagger)=y p_x^5
\end{align*}
where $F(x)=\int {f} dx$ satisfies equation (\ref{d5}).\\
$(c_5,b_1)$:\\
\begin{align*}
V=\hbar ^6 f(\hbar ^2 x)+\frac{\alpha_1}{i \hbar}x,\;K=\alpha_2 K_1-\alpha_1 H_1 K_2
\end{align*}
$(c_5,b_3)$:\\
\begin{align*}
V=\hbar ^6 f(\hbar ^2 x)+\hbar^{2}\omega_{2}^{2}P_{I}(\omega_{2}y),\; \omega_2=\frac{\sqrt[5]{4 i \alpha_2}}{\hbar},\;K=\alpha_2 K_1-\alpha_1 H_1 K_2
\end{align*}
where $f(X)$ satisfies equation (\ref{fif3}).\\
\textbf{Classical superintegrable system:}\\
The systems constructed by $(d_1,d_1), (d_1,d_2)$ and $(d_2,d_2)$, i.e. the Jauch-Hill and Smorodinsky-Winternitz potentials, are the same as in the quantum case. Those related to $(a_i,a_j)$ have no classical analog.  In the approach of this article we generate the following potentials and integrals.\\
$(b_3,b_3):$
\begin{align*}
&V=\epsilon (\sqrt{\frac{2\alpha_1}{3}x}+ \sqrt{\frac{2\alpha_2}{3}y}); \; \epsilon=\pm 1,\\
&K=\alpha_2 K_1-\alpha_1 K_2.
\end{align*}
$(b_1,b_3):$
\begin{align*}
&V=\alpha_1 x+\epsilon \sqrt{\frac{2\alpha_2}{3}y},\\
&K=\alpha_2 K_1-\alpha_1 K_2.
\end{align*}
$(c_2,b_3)$:\\
\begin{align*}
V=\frac{\beta}{x^2}+\epsilon \sqrt{\frac{2\alpha_2}{3}y},\;
K=\alpha_2 K_1-\alpha_1 H_1 K_2
\end{align*}
$(c_5,b_1)$:\\
\begin{align*}
V=f(x)+\alpha_2 y,\;K=\alpha_2 K_1-\alpha_1 H_1 K_2
\end{align*}
$(c_5,b_3)$:\\
\begin{align*}
V=f(x)+\epsilon \sqrt{\frac{2\alpha_2}{3}y},\;K=\alpha_2 K_1-\alpha_1 H_1 K_2
\end{align*}
where $f(x)$ satisfies equation 
$$-(\alpha_1 x-15 f^2) f'=2\alpha_1 f$$
$(c_3,b_1):$
\begin{align*}
&V=g(x)+\alpha_2 y,\;
K=\alpha_2 K_1-\alpha_1 H_1 K_2=\alpha_2 p_x^3-\frac{\alpha_1}{2} p_x^2 p_y.
\end{align*}
$(c_3,c_3)$:\\
\begin{align*}
&V=g(x)+g(y),\; K=\alpha_2 H_2 K_1-\alpha_1 H_1 K_2=(\alpha_2 p_x -\alpha _1 p_y)p_x^2p_y^2.
\end{align*}
$(c_2,c_3)$:\\
\begin{align*}
&V=\frac{\beta}{x^2}+g(y),\; K=\alpha_2 H_2 K_1-\alpha_1 H_1 K_2.
\end{align*}
$(c_3,b_3)$:\\
\begin{align*}
V=g(x)+\epsilon \sqrt{\frac{2\alpha_2}{3}y},\;
K=\alpha_2 K_1-\alpha_1 H_1 K_2
\end{align*}
where $g(x)$ satisfies equation 
$$(\alpha_1 x-2g)^{2}g=c$$
$(d_1,d_3):$
\begin{align*}
&V=\frac{\alpha^{2}}{2} x^{2}+h(y), \alpha_2=\alpha,\; K=(K_1^\dagger)(K_2^{-})-(K_1^{-})(K_2^\dagger)
\end{align*}
$(d_3,d_2):$
\begin{align*}
&V=h(x)+\frac{\alpha ^2 }{8}y^2+\frac{\beta}{y^2},\\
&K=(K_1^\dagger)(K_2^{-})-(K_1^{-})(K_2^\dagger)=(x p_y-y p_x)p_x^2 p_y.
\end{align*}
$(d_1,d_3):$
\begin{align*}
&V=\frac{\alpha^{2}}{2} x^{2}+h(y), \alpha_2=2\alpha\; K=(K_1^\dagger)^2(K_2^{-})-(K_1^{-})^2(K_2^\dagger)=(x p_y-y p_x)p_x p_y^2.
\end{align*}
\begin{align*}
V=\frac{\alpha^{2}}{2}x^{2}+h(y), \alpha_2=3\alpha,\;K=(K_1^\dagger)^3(K_2^{-})-(K_1^{-})^3(K_2^\dagger)=(x p_y -y p_x)p_x^2p_y^2
\end{align*}
$(d_3,d_3)$:\\
\begin{align*}
V=h(x)+h(y),\;K=(K_1^\dagger)(K_2^{-})-(K_1^{-})(K_2^\dagger)=(x p_y-y p_x)p_x^2p_y^2
\end{align*}
where $h$ satisfies the nonlinear ODE (\ref{d3-2}), or equivalently the algebraic equation (\ref{d3-1}).\\
$(d_4,d_1):$
\begin{align*}
&V=k(x)+\frac{\alpha ^2}{2}y^{2},\\
&K=(K_1^\dagger)(K_2^{-})-(K_1^{-})(K_2^\dagger)=(x p_y-y p_x) p_x^3.
\end{align*}
$(d_4,d_2)$:\\
\begin{align*}
V=&k(x)+\frac{\alpha^2}{8}y^2+\frac{\beta}{y^2},\\
K=&(K_1^\dagger)(K_2^{-})-(K_1^{-})(K_2^\dagger)=(x p_y-y p_x) p_x^3 p_y.
\end{align*}
$(d_1,d_4)$:\\
\begin{align*}
V=&\frac{\alpha^{2}}{2}x^{2}+k(y), \alpha_y=2\alpha,\\
K=&(K_1^\dagger)^2(K_2^{-})-(K_1^{-})^2(K_2^\dagger)=(x p_y-y p_x)p_x p_y^3.
\end{align*}
where $k$ satisfies the nonlinear ODE (\ref{d4-2}), or equivalently the fifth order algebraic equation (\ref{d4-1}).\\
$(b_1,b_5)$:\\
\begin{align*}
V=\alpha_1 x+\sqrt[3]{\frac{2\alpha_2}{5}y},\; K=\alpha_2 K_1-\alpha_1 K_2,
\end{align*}
$(b_3,b_5)$:\\
\begin{align*}
&V=\epsilon \sqrt{\frac{2\alpha_1}{3}x}+\sqrt[3]{\frac{2\alpha_2}{5}y}.\\
&K=\alpha_2 K_1-\alpha_1 K_2.
\end{align*}
$(b_5,b_5)$:\\
\begin{align*}
&V=\sqrt[3]{\frac{2\alpha_1}{5}x}+\sqrt[3]{\frac{2\alpha_2}{5}y},\; K=\alpha_2 K_1-\alpha_1 K_2.
\end{align*}
where $f(x)$ and $g(y)$ satisfy equation (\ref{Vb5}).\\
$(d_5,d_1)$:\\
\begin{align*}
&V=f(x)+\frac{\alpha^{2}}{2} y^{2},\;K=(K_1^\dagger)(K_2^{-})-(K_1^{-})(K_2^\dagger)=y p_x^5
\end{align*}
where $F(x)=\int {f} dx$ satisfies equation (\ref{d5-1}).\\\\
\section{Conclusion}

Our main conclusion is that the systematic use of quantum or classical algebraic systems in one dimension is an efficient method of generating superintegrable systems in a two-dimensional Euclidean space. By construction, all systems thus obtained allow the separation of variables in Cartesian coordinates in the Schr\"{o}dinger and the Hamilton-Jacobi equation, respectively. The algebraic systems consist of a pair $(H_1,K_1)$ where $H_1$ is a natural Hamiltonian as in (\ref{H1}) and $K_1$ a polynomial as in (\ref{Lx}). The four types of algebras considered are as in (\ref{types}) and all of them should be  constructed in $x$ and $y$ spaces independently.\\
Let us again run through all combinations of the type $(z_i, w_j)$ where $z$ is in $x$-space and $w$ in $y$-space. The subscripts give the order of the corresponding polynomial $K_l, l=1,2$.\\
The pair $(d_1,d_1)$ with $(\alpha_1, \alpha_2) = \omega^2(n,m)$ yields the Jauch and Hill potentials.\\
The pair $(d_2, d_2)$ gives an infinite set of generalizations of the Smorodinsky-Winternitz potentials in particular the "caged harmonic oscillator" of \cite{RTW08} and \cite{EV08}.\\
Pairs of the type $(a_i,a_j)$ in quantum mechanics give potentials in terms of elliptic or hyper-elliptic functions. In classical mechanics
their limit is free motion ($V=$ constant).\\
All other pairs in quantum mechanics lead to "exotic potentials" expressed in terms of
Painlev\'e transcendents or their generalizations that are solutions of higher order ODEs. This is true for all examples so far considered and we conjecture that this is true for all values of  $i$ and $j$. In the classical case exotic potentials also exist. Very often they satisfy nonlinear algebraic equations. The obtained ODEs are also nonlinear of lower order than in the quantum case and they do not have the Painlev\'e property.



\section*{ACKNOWLEDGEMENTS}

The research of P.W. was partially supported by an NSERC Discovery grant. M.S. thanks the University of Montreal for a "bourse d'admission" and a "bourse de fin d'\'etudes doctorales". The research of I.\ M.\ was supported by the Australian Research Council through Discovery Early Career Researcher Award DE130101067 and
a DP160101376 grant.\par

%


\end{document}